\documentclass[11pt,a4paper]{article}
\usepackage{jheppub}

\usepackage{amsmath, amsfonts, amsthm, amssymb, graphicx}
\usepackage[dvipsnames]{xcolor}
\usepackage{epstopdf}
 \usepackage{slashed}
\usepackage{changepage}
 \usepackage{vector}

\usepackage{pifont}

\usepackage{caption}
\def\be{\begin{equation}}
\def\ee{\end{equation}}
\def\ba{\begin{eqnarray}}
\def\ea{\end{eqnarray}}

\def\bdm{\begin{displaymath}}
\def\edm{\end{displaymath}}

\def\bq{\begin{quote}}
\def\eq{\end{quote}}

\def\del{\partial}
\def\ltap{\ \raise.3ex\hbox{$<$\kern-.75em\lower1ex\hbox{$\sim$}}\ }
\def\gtap{\ \raise.3ex\hbox{$>$\kern-.75em\lower1ex\hbox{$\sim$}}\ }
\def\gl{\ \raise.5ex\hbox{$>$}\kern-.8em\lower.5ex\hbox{$<$}\ }
\def\roughly#1{\raise.3ex\hbox{$#1$\kern-.75em\lower1ex\hbox{$\sim$}}}

 at 10truept

\newcommand{\beq}{\begin{equation}}
\newcommand{\eeq}{\end{equation}}
\newcommand{\bea}{\begin{eqnarray}}
\newcommand{\eea}{\end{eqnarray}}
\newcommand{\beqa}{\begin{eqnarray}}
\newcommand{\eeqa}{\end{eqnarray}}

\def \del {\partial}

\def\be{\begin{equation}}
\def\ee{\end{equation}}

\usepackage[dvipsnames]{xcolor}

\title{Vacuum transitions with the Gauss-Bonnet term in $D$ dimensions}

\author[1,2]{Yang Liu,}

\affiliation[1]{School of Physics and Astronomy, University of Nottingham, University Park, Nottingham NG7 2RD, United Kingdom}
\affiliation[2]{Nottingham Centre of Gravity, University of Nottingham, Nottingham NG7 2RD, UK}

\emailAdd{yang.liu@nottingham.ac.uk}

\abstract{In our previous paper \cite{us,LPP2024}, we proposed a probabilistic argument to explain the reason why the cosmological constant is very small in $4D$. We can ask a question: if the behavior of tunneling exponent $B$ can be generalized to $D$-dimension. Moreover, in higher dimensional theory motivated by string theory the Gauss-Bonnet term plays an important role. Therefore, in this paper, we generalize our result in \cite{us,LPP2024} to arbitrary $D$ dimensions including the Gauss-Bonnet term. As a result, we have two main results. We find that the Euclidean action of the bounce, $B$, describing the decay of a de Sitter vacuum, is proportional to $k^{-(D-2)}_{+}$, which has a pole as $k^2_{+} \rightarrow 0$ where $k^2_{+}$ is the curvature of the parent vacuum. This result is similar to the result in $4D$. The other result is that we find a new decay channel, describing up-tunneling from anti-de Sitter into de Sitter. The meaning of this new decay channel in the string landscape should be explored in the future. } 

\begin{document}

\maketitle

\section{Introduction}
One of the most important discoveries in modern cosmology is that our universe is undergoing accelerated expansion, which is driven by the so-called “dark energy”. The simplest explanation of dark energy is that it is the vacuum energy which gravitates like a cosmological constant. One of the astonishing features of the cosmological constant is that its value is extremely small and remains constant over time.

How to explain the origin of the small value of the cosmological constant has become a long standing problem in theoretical physics. Pauli did the calculation of he zero-point energy of the electron on the curvature of spacetime and he found that the universe cannot reach to the moon \cite{Enz}. According to quantum field theory, we know that when we compute the contributions to the vacuum energy which are from quantum fields, the radiative corrections to the vacuum energy are extremely sensitive to the UV cut-off. Radiative corrections scale like the 4-th power of the UV cut-off. For example, if we take the cut-off to TeV scale or beyond, then the vacuum energy density of the theoretical result lies at least 60 orders of magnitude higher than the scale of dark energy. Furthermore, if we take the UV cut-off to the Planck scale, then the discrepancy between theory and observation will extend to 120 orders of magnitude. More reviews of the cosmological constant problem can be found in \cite{Weinberg:1988cp, Polchinski:2006gy, Burgess:2013ara,Padilla:2015aaa}.
 
In recent two papers, Kaloper \cite{Kaloper:2022oqv} and Kaloper and Westphal \cite{Kaloper:2022jpv} constructed a simple model where the cosmological constant can be neutralized by membrane nucleation. The membranes are charged under a pair of 3-forms. The authors can obtain a dense vacua landscape which includes the current vacuum by introducing the so-called “irrational trick”. One important thing of their models is that the membrane charges are not very small in Planck units. Phase transitions between two different vacua go through via the nucleation of membranes. By assuming some relatively mild assumptions on the membrane charge and tension, the authors obtain that the nucleation tunneling rate slows down exponentially quickly. The vacua decay stops when curvature goes to zero. In fact, the model of Kaloper and Westphal is a generalization of covariant formulation of unimodular gravity of Henneaux and Teitelboim \cite{Henneaux:1989zc}. However, although it is phenomenologically interesting, the model is still challenging to be embedded within fundamental theory such as string theory since the model has no kinetic term.

However, in our previous paper we have shown that the key aspects of the neutralization mechanism of the model can be applied to a general four-dimensional effective field theory \cite{us}. The famous Bousso-Polchinski set-up \cite{Bousso:2000xa} is one of the examples of the 4D EFT. In current string cosmology, anthropic principle has become the dominant explanation for the small value of cosmological constant. We proposed a “probabilistic” argument for the cosmological constant to challenge the anthropic principle. The general family of EFTs included multiple species of four-forms coupled to multiple scalars with generic potentials. We can hope that the relaxation mechanism can be derived from string flux compactifications. In our model the three-forms and the scalars gravitate like a cosmological constant in vacuum, even in the presence of a four-form flux. The four-form flux is quantised, which gives rise to a landscape of vacua \cite{Bousso:2000xa}. In order to match with the observations, the resulting landscape must be so dense near Minkowski spacetime that we can get a very small cosmological constant.  

We considered the cosmological constant problem in our previous paper \cite{us}. In the paper we computed the tunnelling exponent $B$,
\be \label{B}
B=S_E(\text{instanton})-S_E(\text{parent}),
\ee
where $S_E(\text{instanton})$ is the Euclidean action of instanton and $S_E(\text{parent})$ is the Euclidean action of parent vacuum. We will explain more details about $B$ in the following.

The cosmological constant problem is of course a $4D$ problem. We have explored the transition between vacua in $4D$ \cite{us,LPP2024}. If we want to consider the properties of string landscape in higher dimensions, we can ask if  exist transition between vacua in higher dimensions. Furthermore, the physics in higher dimensions can affect the physics in $4D$ after string compactification \cite{Polchinski:1998rr}. Moreover, in higher dimensional theory motivated by string theory the Gauss-Bonnet term plays an important role \cite{CELP2022}. It is usually considered that the Gauss-Bonnet term has no dynamical effect in $4D$ since it is a topological term in $4D$. However, in \cite{GL2020} the authors found that in the limit of strong coupling, the Gauss-Bonnet term in $4D$ gives rise to non-trivial contributions to gravitational dynamics, while preserving the number of graviton degrees of freedom and being free from Ostrogradsky instability. Therefore, we can ask if the Gauss-Bonnet term could play a role in solving the cosmological constant problem.

So in the present paper, we will study the transition between vacua, or in other words, the tunneling rates of vacuum decay, in higher dimensions including the Gauss-Bonnet term. According to the stability analysis of vacua in Gauss-Bonnet gravity, the associated coupling constant of the Gauss-Bonnet term, $\theta$, should be positive, otherwise the whole theory will be unstable \cite{CP2008}. In the following calculation, we assume that $\theta$ is positive. We should point out that there exist two possibilities if we require positive membrane tension. We can find a new decay branch $AdS_{+} \rightarrow dS_{-}$ in one of the possibilities. This new branch does not exist in General Relativity and our previous model \cite{us}. We have not figured out the meaning of the new decay branch in the string landscape so far. It should be studied in the future.

The rest of this paper is organised as follows: in section 2, we present the generalised set-up and the landscape of Lorentzian vacuum solutions. In section 3, we Wick rotate to Euclidean signature and solve the Euclidean field equations to find the corresponding instanton solutions and compute the corresponding transition rates. We demonstrate the role of the parameter $X$ in controlling the stability of near Minkowski vacua, protecting them from decay into anti-de Sitter. We find that if we consider the effect of the Gauss-Bonnet term, there exists a new decay branch, $AdS_{+}$ to $dS_{-}$, which does not exist in General Relativity. In section 4, we discuss our results.

\section{The generalised set-up and vacua} \label{sec:setup}

\subsection{The generalised set-up }
We start with a general $D$-dimensional effective theory on a manifold, $\mathcal{M}$, with a dynamical metric $g_{\mu\nu}$ and a family of $(D-1)$-form fields, $A^i$,  and dual scalars $\phi_i$, 
\begin{multline} \label{Lorentzianaction}
S =\int_\mathcal{M} d^D x\sqrt{|g|} \left[\frac{1}{2}M_{pl}^{D-2}  R-\frac12 \omega^{ij}(\phi) \nabla_\mu \phi_i \nabla^\mu  \phi_j-V(\phi)+ \theta(\phi) R_{GB} \right] \\ +
\int_\mathcal{M} \left[ -\frac{1}{2}Z_{ij} (\phi)F^i\wedge \star F^j+\sigma_i(\phi) F^i \right]
 +S_\text{boundary}+S_\text{membranes} ,
\end{multline}
      where $D$ is the dimension of spacetime and $R$ the Ricci scalar. $\theta$ is a linear Gauss-Bonnet coupling parameter, which is a function of scalar fields and $R_{GB}=R_{\mu\nu\rho\sigma} R^{\mu\nu\rho\sigma}- 4R_{\mu\nu}R^{\mu\nu} + R^2$ is the associated Gauss-Bonnet term. The $D$-form field strengths $F^i$ are given in terms of the $(D-1)$-form fields $F^i=dA^i$ and $\star$ denotes the Hodge star operator on the manifold $\mathcal{M}$.  
      
The action is also equipped with boundary terms which depend on the choice of boundary conditions. These are integrals over the boundary, $\Sigma$, which we take to be a co-dimension one surface described by the embedding $x^\mu=X^\mu(\xi^a)$. The induced metric and the pullback of the $(D-1)$-forms on the boundary are given respectively by $\gamma_{ab}=g_{\mu\nu}X^\mu_{, a}X^\nu_{, b}$ and $\alpha^i=\frac{1}{(D-1)!} A^i_{\mu_1 \mu_2 ... \mu_{D-1}} X^{\mu_1}_{, i_1}X^{\mu_2}_{, i_2} ...X^{\mu_{D-1}}_{, i_{D-1}} d\xi^{i_1} \wedge d \xi^{i_2} ... \wedge d \xi^{i_{D-1}}$, where $X^\mu_{, a}=\del X^\mu/\del \xi^a$ are the boundary tangent vectors. For an action of the form \eqref{Lorentzianaction}, the analogue of the Gibbons-Hawking boundary term, is given by \cite{PS2012,Davis2003,CLP2019},
\begin{equation} \label{Lorentzianboundary}
\int_{\Sigma} d^{D-1}x \sqrt{|\gamma|} \left[M^{D-2}_{pl} K+ 4\theta (J-2\hat{G}^{ij} K_{ij}) \right] -\int_{\Sigma} \mu p^i \phi_i  +\lambda \chi_i\alpha^i,
\end{equation}
where we define ``conjugate momenta", 
\be
p^i=-d^{D-1} \xi \sqrt{|\gamma|} \omega^{ij} n^\mu \nabla_\mu \phi_j,\quad \chi_i= \sigma_i-Z_{ij}( \star F^j). 
\ee
In eq.\eqref{Lorentzianboundary}, $\gamma_{ij}$ is the induced metric on the spacetime boundary, $\Sigma$, with corresponding Einstein tensor $\hat{G}^{ij}$. The extrinsic curvature, $K_{ij} = \frac{1}{2} \mathcal{L}_n \gamma_{ij}$, is defined in terms of the Lie derivative of the induced metric with respect to the $outward$ pointing normal, $n^a$, and $K=\gamma^{ij}K_{ij}$ is its trace. Finally, we can define \cite{Davis2003,CLP2019}
\begin{equation} \label{Jij}
J_{ij} = \frac{1}{3} \left[(K_{kl}K^{kl} - K^2)K_{ij} + 2K K_{ik} K^k_{\ j} - 2 K_{ik} K^{kl} K_{lj}  \right]
\end{equation}
along with its trace $J=\gamma^{ij}J_{ij}$. The extrinsic curvature piece ensures that the action can be extremised under metric variations with Dirichlet boundary conditions, $\delta \gamma_{ab}=0$ \cite{Gibbons:1976ue}. Meanwhile, the parameter $\lambda$ allows us to interpolate between Dirichlet ($\lambda=0$) and Neumann ($\lambda=1$) boundary conditions on the $(D-1)$-forms, while the parameter $\mu$ allows us to interpolate between Dirichlet ($\mu=0$) and Neumann ($\mu=1$)  boundary conditions on the scalars \cite{us}. 

Moreover, we consider the membrane contributions. We can include contributions from membranes and anti-membranes, $\Sigma_I$,  charged under any of the $(D-1)$-forms, such that
\be \label{membact}
S_\text{membranes}=-\sum_I \left\{\eta^i_I q_i\ \int_{\Sigma_I} \alpha_I^i + \tau_i \int_{\Sigma_I} d^{D-1} \xi \sqrt{|\gamma_I}| \right\}.
\ee
Membranes charged under $A^i$ carry a fundamental charge $\pm q_i$ depending on whether they are branes or antibranes and tension $\tau_i$. In the action \eqref{membact},  $\eta^i_I=0, \pm 1$ depending on whether the membrane $\Sigma_I$ carries positive ($\eta^i_I= 1$), negative ($\eta^i_I=- 1$) or vanishing charge ($\eta^i_I=0$) under $A^i$.  The pullback and the induced metric of the three-forms on $\Sigma_I$ are given in a similar way to the boundary, by $\alpha^i_I=\frac{1}{(D-1)!} A^i_{\mu_1 \mu_2 ... \mu_{D-1}} X^{\mu_1}_{I, i_1}X^{\mu_2}_{I, i_2} ...X^{\mu_{D-1}}_{I, i_{D-1}} d\xi^{i_1} \wedge d \xi^{i_2} ... \wedge d \xi^{i_{D-1}}$ and $\gamma_I{}_{ab}=g_{\mu\nu}X_I{}^\mu_{, a}X_I{}^\nu_{, b}$, where $X_I{}^\mu_{, a}=\del X_I{}^\mu/\del \xi^a$ are the  tangent vectors on $\Sigma_I$. In the current paper we only consider the timelike membranes so that their unit normal $n^{\mu}_I$ is spacelike.

\subsection{Vacua}
We take vacua to be real Lorentzian solutions with constant scalars, $D$-forms of constant flux $\star F^i=c^i$, and a maximally symmetric metric with constant curvature $k^2$, corresponding to dS ($k^2>0$),  Minkowski ($k^2=0$) or AdS ($k^2<0$) spacetime. Since the scalars are constant, then the Gauss-Bonnet coupling parameter $\theta$ is a constant as well. The conjugate momentum, 
\be
\chi_i = \sigma_i- Z_{ij} c^j  \label{con3}
\ee
is locally constant. The equations of motion away from membranes will be given in \eqref{Neq}, \eqref{rhoeq} and \eqref{Aeq}. At membranes, we get a jump in $\chi_i$,
\be \label{chijump}
\Delta \chi_i = -\eta^i_I q_i \qquad  (\text{no sum over $i$}),
\ee
generically triggering a jump in the spacetime curvature. Thus we have a landscape of possible vacua with different cosmological constants, scanned through membrane nucleation. Membrane nucleation, which is a quantum process, is necessary for scanning the landscape of vacua.  When we compute transition rates we do so between eigenstates of constant $\chi_i$. This suggests a path integral formalism equipped with Neumann boundary conditions, fixing $\chi_i$ in both the in-state and in the out-state \cite{Duncan:1989ug}.  We will compute these transition rates in the next section. 

\section{Nucleation rates} \label{sec:rates}
Transitions between vacua go through via membrane nucleation. In order to compute the nucleation rate at which transitions occur, we need to do analytical continuation to Euclidean signature
\be
t \to -i t_E, \qquad \star F^i \to \star F^i, \qquad A^i\to iA^i, \qquad S \to iS_E,
\ee
where $S_E$ is the Euclidean action 
\begin{multline} \label{euclideanaction}
S_E =-\int_\mathcal{M} d^D x_E\sqrt{|g|} \left[\frac{1}{2}M_{pl}^{D-2}  R-\frac12 \omega^{ij}(\phi) \nabla_\mu \phi_i \nabla^\mu  \phi_j-V(\phi) + \theta R_{GB} \right] 
\\+\int_\mathcal{M} \left[ -\frac{1}{2}Z_{ij} (\phi)F^i\wedge \star F^j+\sigma_i(\phi) F^i \right]
 +S^E_\text{boundary}+S^E_\text{membranes},
\end{multline}
Moreover, we should solve for the instanton solution interpolating between the parent vacuum, $\mathcal{M}_+$, with curvature $k_+^2$ and the daughter vacuum, $\mathcal{M}_-$, with curvature $k_-^2$.\\
The boundary terms have been chosen to be consistent with Neumann boundary conditons on the $(D-1)$-form fields ($\lambda=1$). We will only consider bounce configurations that transition between two vacua, therefore we assume that there is a single membrane, $\Sigma$, which is charged under $A^i$ for some particular choice of $i=i_*$. The membrane tension is $T=\tau_{i_*}$ and the membrane charges under $A^i$ are given by $Q_{i}=\delta_{i i_*} Q_{i_*}$, where $Q_{i_*}=\pm q_{i_*}$. After the Wick rotation, the Euclidean action of membrane is given by
 \be
 S^E_\text{membranes}=- Q_{i_*}\ \int_{\Sigma} \alpha_\Sigma^{i_*} + T \int_{\Sigma} d^{D-1} \xi_E \sqrt{| \gamma_{\Sigma} |}\, .
 \ee\\
We consider $O(D)$ symmetric Euclidean field configurations, with metric
\be
ds^2=N^2(r) dr^2+\rho(r)^2 d \Omega_{D-1},
\ee
where $d \Omega_{D-1}=h_{ij} d\xi^i d\xi^j$ is the metric on a unit $(D-1)$-sphere, Euclidean $(D-1)$-form potentials 
\be
A^i=A^i(r) \sqrt{|h|} d^{D-1} \xi
\ee
and scalars $\phi_i=\phi_i(r)$. We will set $N(r)=1$ in the following calculation. The membrane is assumed to lie at $r=0$ and radial coordinate runs from $r_\text{min}<0$ to $r_\text{max}>0$.  
Note that upon Wick rotation back to the Lorentzian signature, the instanton solution corresponds to a bubble of daughter vacuum in the parent spacetime. 

With this ansatz, the resulting field equations  away from the membrane for constant scalars $\phi_i$ are the following (we have set $N(r)=1$):\\
For $N$:
\begin{equation} \label{Neq}
\begin{split}
 0= &\frac{1}{2}(D-1)(D-2) M_{pl}^{D-2} \left[\frac{1}{\rho^2} -\left(\frac{\rho'}{\rho}\right)^2 \right]  - V-\frac12 Z_{ij} \frac{A^i{}' A^j{}'}{\rho^{2D-2}}\\
 & + \theta (D-1)(D-2)(D-3)(D-4) \frac{1}{\rho^4} (1-\rho'^2)^2 
 \end{split}
\end{equation}
For $\rho$:
\begin{equation} \label{rhoeq}
\begin{split}
 0= & \frac{1}{2} (D-1)(D-2) M_{pl}^{D-2}\left[\frac{D-3}{\rho^2} - (D-3)\left(\frac{\rho'}{\rho}\right)^2 -2\frac{\rho''}{\rho}\right]\\
  & -(D-1)V -\frac{D-1}{2} Z_{ij} \frac{A^i{}' A^j{}'}{\rho^{2D-2}}\\ 
  & + \theta (D-1)(D-2)(D-3)(D-4)(D-5) \frac{1}{\rho^4} (1-\rho'^2)^2\\
  & + 4\theta (D-1)(D-2)(D-3)(D-4) \frac{\rho'^2}{\rho^3} \rho''\\
  & - 4\theta (D-1)(D-2)(D-3)(D-4) \frac{\rho''}{\rho^3} 
\end{split}
\end{equation}
For $A^i$:
\begin{equation} \label{Aeq}
  \chi_i'=0 
\end{equation}
where $\chi_i=\sigma_i-Z_{ij} \frac{A^j{}'}{\rho^{D-1}}$ and $'$ denotes the derivative with respect to $r$.

Since we have assumed that all scalars are locally constant and satisfy the constraints listed in \eqref{rhoeq} and \eqref{Aeq}, this system is solved by
\be \label{rho}
\rho=\frac{\sin(k(\epsilon r+r_0)}{k}, \qquad \epsilon=\pm 1
\ee
where $k^2$ is the curvature of spacetime, and $(D-1)$-form potentials
\be
A^i(r)=A^i(0)+c^i\int_0^r dr \rho(r)^{D-1}. \label{Asol}
\ee
Note that the solution for $\rho$ extends to $k^2 \leq 0$ by analytic continuation.  The parameter $r_0$ is an integration constant, setting the radius of the $(D-1)$-sphere at the membrane. 

For $k^2>0$, the geometry is that of a section of a $D$-sphere and we can check that under $\epsilon \to -\epsilon$, $r_0 \to \pi/k-r_0$, $\rho$ is invariant. Therefore, without loss of generality, we can always set $\epsilon=+1$, while also assuming $r_0 \in [0, \pi/k]$. The poles of the $D$-sphere are located at $r_\text{max}=\frac{\pi}{k}-r_0$ and $r_\text{min}=-r_0$.

For $k^2=0$ and $k^2<0$, the geometries are different. The geometry for $k^2=0$ is that of a section of a $D$-dimensional Euclidean plane. While the geometry for $k^2<0$ is that of a section of a $D$-dimensional hyperboloid. We can take $r_0\geq 0$ since the $\rho$ at the membrane is non-negative. Unlike the case of $k^2>0$, for these two cases we cannot always set $\epsilon$ without loss of generality and must consider each case separately. For $\epsilon=-1$, $r_\text{max}=r_0$ corresponding to where the $(D-1)$-spheres shrink to zero size, while $r_\text{min}=-\infty$ corresponding to the point where they diverge. For $\epsilon=+1$, the result is reversed: $r_\text{max}=\infty$ corresponding to where the $(D-1)$-spheres diverge, while $r_\text{min}=-r_0$ corresponding to the point where they shrink to zero size.  

We have a continuity constraint at the membrane, 
\be
\Delta \rho(0) =\Delta \left[ \frac{\sin k r_0}{k} \right]=0\, ,
\ee
where we introduce two notations $\Delta x=x^+-x^-$ and $\langle x\rangle=\frac12 (x^++x^-)$, which correspond  to the difference and average of some quantity $x$ defined on either side of the membrane, respectively.

We also have the following junction conditions
\begin{equation} \label{rhojunc}
\begin{split}
 T=  & -(D-2) M_{pl}^{D-2} \Delta \left[ \frac{\rho'(0)}{\rho(0)}\right] +\frac{4}{3} \theta (D-2) (D-3) (D-4) \Delta \left[ \frac{\rho'^3(0)}{\rho^3(0)}\right]  \\
   &- 4 \theta (D-2) (D-3) (D-4) \Delta \left[ \frac{\rho'(0)}{\rho^3(0)}\right] ,   
\end{split}
\end{equation}
\begin{equation} \label{chijunc}
    \Delta \chi_i = -\delta_{i i_*} Q_{i_*}.
\end{equation}
According to eq.\eqref{rhojunc}, the Gauss-Bonnet term has no contribution when $D \leq 4$.

Physically realistic  membranes  always carry non-negative tension, resulting in the following constraint on the allowed configurations
\be 
A \Delta \rho'(0) >0, \label{Tg0}
\ee
where 
\begin{equation} \label{A}
\begin{split}
 A=  & - M_{pl}^{D-2} \rho^2(0) +\frac{8}{3} \theta (D-3) (D-4) \langle \rho'^2(0)\rangle \\
   &- \frac{1}{3} \theta (D-3) (D-4) \Delta \rho'^2(0) - 4\theta (D-3) (D-4).   
\end{split}
\end{equation}
If $\theta$ is zero and $D=4$, then we have 
\begin{equation} \label{the0D4}
    -M^2_{pl} \rho^2(0) \Delta \rho'(0) >0,
\end{equation}
namely,
\be
\Delta \left[ \epsilon \cos k r_0 \right] \leq 0\, , \label{Tcon}
\ee
which is the result we have obtained in \cite{us}.  

Eq. \eqref{A} has two possibilities when $D \geq 5$. The first possibility is $A<0$. Then we have $\Delta \rho'(0) <0$, namely,
\begin{equation} \label{Al0}
    \Delta \left[ \epsilon \cos k r_0 \right] \leq 0\, .
\end{equation}
Then this possibility has the same allowed configurations as we obtained in our previous paper \cite{us}. For this possibility, we have only three configurations that avoid any problems with negative brane tension or infinitely suppressed tunnelling rates. These are the configurations of physical interest corresponding to
\begin{itemize}
    \item $\text{dS}_+ \to \text{dS}_-$
    \item $\text{dS}_+ \to \text{Minkowski/AdS}_-$
    \item $\text{Minkowski/AdS}_+ \to \text{Minkowski/AdS}_-$  $(|k_-| \geq |k_+|)$
\end{itemize}
each with $\epsilon_\pm=1$ and so $\rho'(r_\text{min})=1$. Table 1 summaries all possible transformations for this possibility \cite{us}. \\
\begin{center}
\begin{table}[t]
\centering 
\resizebox{1.1\textwidth}{!}{%
\begin{tabular}{ |c||c|c|c|}
\hline
& $\text{dS}_+$& $\text{Minkowski/AdS}_+$ & $\text{Minkowski/AdS}_+$\\
& $ \epsilon_+=+1$&  $\epsilon_+=+1$ & $\epsilon_+=-1$\\
\hline
\hline
\begin{tabular}[t]{c}$\text{dS}_-$ \\  $\epsilon_-=+1$ \end{tabular}
& 
\begin{tabular}[t]{@{}c|c@{}}
         $(kr_0)_+ \geq \frac{\pi}{2} \geq (k r_0)_-$ & 
         $(kr_0)_+ \geq (k r_0)_-  \geq \frac{\pi}{2}$ \\
         allowed for $|X| \leq 1$  &
          allowed for $X \leq -1$
          \\ \hline
           $ \frac{\pi}{2} \geq (kr_0)_+ \geq (k r_0)_-$   & $(kr_0)_+ < (k r_0)_-$  \\ 
          allowed for $X \geq 1$  & 
         negative tension
      \end{tabular}
& negative tension &
 \begin{tabular}[t]{@{}c|c@{}}  
         $(kr_0)_- \geq \frac{\pi}{2}$ & 
         $\frac{\pi}{2} \geq  (kr_0)_-  $ \\
         kinematically & kinematically \\
         allowed for $X \leq -1$,  &
          allowed for $-1 \leq X \leq 0$, \\
           infinitely suppressed &  infinitely suppressed
            \end{tabular}
  \\
 \hline
 \begin{tabular}[t]{c}$\text{Minkowski/AdS}_-$\\ $\epsilon_-=+1$ \end{tabular}&  \begin{tabular}[t]{@{}c|c@{}}
         $(kr_0)_+ \geq \frac{\pi}{2}$ & 
         $\frac{\pi}{2} \geq  (kr_0)_+ $ \\
allowed for $0\leq X \leq 1$  &
~~allowed for $X \geq 1 $~~~
            \end{tabular} &
 \begin{tabular}[t]{@{}c|c@{}}  
         $|k_-| \geq |k_+|$ & 
         $|k_-| < |k_+|$ \\
         allowed for $X \geq 1$ &
         negative tension
            \end{tabular}
 &  \begin{tabular}[t]{c} kinematically allowed for $|X| \leq 1$,\\ infinitely suppressed \end{tabular} \\ \hline
\begin{tabular}[t]{c} $\text{Minkowski/AdS}_-$\\$\epsilon_-=-1$ \end{tabular} &  negative tension  & negative tension & \begin{tabular}[t]{@{}c|c@{}}  
         $|k_-| > |k_+|$ & 
         $|k_-| \leq |k_+|$ \\
         negative tension
          &  kinematically \\ & ~~allowed for $X \leq -1$, \\ &~~ infinitely suppressed
            \end{tabular}
 \\
 \hline
\end{tabular}%
}
\caption{Summary of transitions $\mathcal{M}_+ \to \mathcal{M}_-$ where $\mathcal{M}$ is de Sitter, Minkowski or anti de Sitter, are listed. The transition is forbidden if a transition has negative tension and we mark it accordingly. We mark whether they are kinematically allowed for $|X|\geq 1$ or $|X| \leq 1$ for the remaining transitions. In some of these cases, we rule out some transitions although they are kinematically allowed since the transition rate is infinitely suppressed. We can take $\epsilon=+1$ for all dS configurations without loss of generality, so any examples contrary to this are not applicable. While for AdS and Minkowski configurations, we need to consider the two cases $\epsilon = +1$ and $\epsilon =-1$, respectively. This table is given in \cite{us}.} \label{tab1}
\end{table}
\end{center}
The second possibility is $A>0$. Then we have $\Delta \rho'(0) >0$, namely,
\begin{equation} \label{Ag0}
    \Delta \left[ \epsilon \cos k r_0 \right] \geq 0\, .
\end{equation}
This possibility has some new allowed configurations. For the second possibility, we also have three configurations that avoid any problems with negative brane tension or infinitely suppressed tunnelling rates as well. These are the configurations of physical interest corresponding to
\begin{itemize}
    \item $\text{dS}_+ \to \text{dS}_-$
    \item $\text{Minkowski/AdS}_+ \to \text{dS}_-$
    \item $\text{Minkowski/AdS}_+ \to \text{Minkowski/AdS}_-$  $(|k_+| \geq |k_-|)$
\end{itemize}
where the second configuration, $\text{Minkowski/AdS}_+ \to \text{dS}_-$, is a new configuration, which is not possible in General Relativity and our previous paper \cite{us}. Table 2 summaries all possible transformations for this possibility.\\
It is not hard to get positive $A$. For example, if we take $\theta = 0.8M^{D-2}_{pl}$, $r_0=0.5$, $|k_+|=0.8$ and $\cos{(k_{-} r_0)}=0.9$, then we can have $A>0$. We define $\theta_{eff} = \theta (D-3)(D-4)$. Since the stability of Gauss-Bonnet gravity requires $\theta$ to be positive \cite{CP2008}, considering $A>0$, we can have the sufficient condition which $\theta$ should satisfy
\begin{equation} \label{thetaeff}
    \theta_{eff} > \frac{M^{D-2}_{pl} \frac{1}{|k^2_{+}|} [\cosh^2(|k_+|r_0)-1]}{\Sigma},
\end{equation}
where 
\begin{equation} \label{Sigma}
    \Sigma = \frac{1}{3} \cosh^2 (|k_+|r_0) + \frac{1}{3} \cos^2{(k_{-} r_0)} + \frac{1}{3} \cosh (|k_+|r_0) \cos{(k_{-} r_0)}-4.
\end{equation}

\begin{center}
\begin{table}[t]
\centering 
\resizebox{1.1\textwidth}{!}{%
\begin{tabular}{ |c||c|c|c|}
\hline
& $\text{dS}_+$& $\text{Minkowski/AdS}_+$ & $\text{Minkowski/AdS}_+$\\
& $ \epsilon_+=+1$&  $\epsilon_+=+1$ & $\epsilon_+=-1$\\
\hline
\hline
\begin{tabular}[t]{c}$\text{dS}_-$ \\  $\epsilon_-=+1$ \end{tabular}
& 
\begin{tabular}[t]{@{}c|c@{}}
         $(kr_0)_- \geq \frac{\pi}{2} \geq (k r_0)_+$ & 
         $(kr_0)_- \geq (k r_0)_+  \geq \frac{\pi}{2}$ \\
         allowed for $|X| \leq 1$  &
          allowed for $X \geq 1$
          \\ \hline
           $ \frac{\pi}{2} \geq (kr_0)_- \geq (k r_0)_+$   & $(kr_0)_+ > (k r_0)_-$  \\ 
          allowed for $X \leq -1$  & 
          negative tension 
          \end{tabular} &
           \begin{tabular}[t]{@{}c|c@{}}  
         $(kr_0)_- \leq \frac{\pi}{2} $ & 
         $(kr_0)_- > \frac{\pi}{2} $ \\
         kinetically & kinetically \\
         ~~allowed for $X \leq -1$ & ~~allowed for $-1 < X <0 $ \\
         &
         \end{tabular} &
 negative tension 
  \\
 \hline
 \begin{tabular}[t]{c}$\text{Minkowski/AdS}_-$\\ $\epsilon_-=+1$ \end{tabular}&  negative tension &
  \begin{tabular}[t]{@{}c|c@{}}
         $|k_+| \geq |k_-|$ & 
         $|k_+| < |k_-|$ \\
         kinetically & negative tension \\
allowed for $X < -1$  & 
            \end{tabular} &
negative tension \\ \hline
\begin{tabular}[t]{c} $\text{Minkowski/AdS}_-$\\$\epsilon_-=-1$ \end{tabular}  & \begin{tabular}[t]{@{}c|c@{}}  
         $(k r_0)_+ \leq \frac{\pi}{2} $ & $(k r_0)_+ \geq \frac{\pi}{2}$ \\
         kinetically & kinetically \\
        allowed for $0 \leq X \leq 1$  & allowed for $X \geq 1$ \\
        infinitely suppressed & infinitely suppressed
            \end{tabular}
&  \begin{tabular}[t]{c} kinematically \\ allowed for $|X| \leq 1$,\\ infinitely suppressed \end{tabular} 
& \begin{tabular}[t]{@{}c|c@{}}
         $|k_+| \leq |k_-|$ & $|k_+| > |k_-|$ \\
          kinetically & negative tension \\
          allowed for $X \leq -1$  &  \\
          infinitely suppressed &
            \end{tabular} 
 \\
 \hline
\end{tabular}%
}
\caption{Summary of transitions $\mathcal{M}_+ \to \mathcal{M}_-$ when $A>0$. } \label{tab2}
\end{table}
\end{center}

The tunnelling rates between two vacua $\mathcal{M}_+ \to \mathcal{M}_-$ in semi-classical theory of vacuum decay are given by the following formula
\be
\frac{\Gamma}{\text{Vol}} \sim e^{-B/\hbar},
\ee
where the tunnelling exponent $B$ is 
\be
B=S_E(\text{instanton})-S_E(\text{parent}).
\ee
and $\Gamma$ is the transition (or tunnelling) rate. Here $S_E(\text{instanton})$ is the Euclidean action evaluated on the bubble configurations described above, interpolating between the two vacua $\mathcal{M}_+$ and $\mathcal{M}_-$.  While $S_E(\text{parent})$ is the Euclidean action evaluated on the complete parent vacuum,  $\mathcal{M}_+$, which has no bubbles.

After a relatively lengthy calculation and plenty of heart warming cancellations, the tunnelling exponent $B$ can be computed in all cases:
\be\label{bounce}
\begin{split}
  \frac{B}{\Omega_{D-1}}=\Delta & [ \int^0_{r_\text{min}} dr \frac{(D-1)(D-2)}{2} M^{D-2}_{pl} \frac{1}{k^{D-3}} \left(1+\cos^2 k(r_0 +r) \right) \sin^{D-3} k(r_0 +r)  \\ 
  & +\theta (D-1)(D-2)(D-3)(D-4) \frac{1}{k^{D-5}} \sin^{D-5} k(r_0 +r) \\
  & \times \left[1- \frac{1}{3} \cos^4 k(r_0 +r) +2 \cos^2 k(r_0 +r) \right]\\
  & +\theta (D-1)(D-2)(D-3)(D-4) \frac{1}{k^{D-5}} \sin^{D-1} k(r_0 +r) \\
  & - \frac{(D-1)(D-2)}{2} M^{D-2}_{pl} \frac{1}{k^{D-3}} \sin^{D-1} k(r_0 +r) ] + T\rho(0)^{D-1} ,  
\end{split}
\end{equation} 
where $r_\text{min}$ denotes the minimal value of the radial coordinate and  $\Omega_{D-1}$ is the volume of the unit $(D-1)$-sphere. In \eqref{bounce}, we have used \eqref{rho} for $dS$ vacua. Of course, for $AdS$ vacua, $\rho$ should be replaced by
\be \label{rhoAdS}
\rho_{AdS} = \frac{1}{|k|} \sinh (|k| (\epsilon r + r_0)), \quad \epsilon= \pm 1.
\ee

As we did in \cite{us}, we now define a parameter $X$, which is given by
\be \label{Xz}
X=-\frac{2\langle \rho'(0) \rangle}{ \Delta [\rho'(0)]} \equiv \frac{1+z}{1-z}, \qquad z\equiv \frac{(\epsilon \cos kr_0)_+}{(\epsilon \cos kr_0)_-}.
\ee
The value of $X$ now corresponds to a kinematic constraint on the geometry. When $z<1$, $X \in (-1, +\infty)$; when $z>1$, $X \in (-\infty, -1)$. Since the values of $X$ in these two parts are not continuous, we can expect that the behaviour near $X=1$ should be very important. 

According to the junction conditions \eqref{rhojunc} and \eqref{chijunc}, $X$ can also be rewritten in terms of $\Delta k^2$ and $T^2$. The expression is complicated and we do not use it in the present paper, thus we do not list the formula for $X$. In particular, if $\theta=0$, we can get
\be \label{XDnoGB}
X=\frac{(D-2)^2 M_{pl}^{2(D-2)} \Delta k^2}{T^2}
\ee
with $X>0$ in downward transitions and $X<0$ in upward transitions. Furthermore, if $D=4$, we have $X= \frac{4 M^4_{pl} \Delta k^2}{T^2}$, which is consistent with the $X$ we have obtained in \cite{us}. More importantly, however, it turns out that some configurations are kinematically allowed only when $|X|\leq 1$ with the remainder allowed when $|X|\geq 1$. 

When we consider the down-tunnelling from near Minkowski space to anti-de Sitter space, which $k_{+}$ goes to zero while $k_{-}$ is finite, according to \eqref{bounce}, the tunnelling exponent for $Min_{+} \rightarrow AdS_{-}$ goes as  
\be
B_{M_+ \to AdS_-} \sim \frac{1}{k_+^{D-2}},\qquad X \approx 1,
\ee
where $k^2_+$ is the curvature of parent vacuum. As the parent vacuum approaches Minkowski, this exponent diverges, suppressing any transition into AdS and ensuring a long lived Minkowski vacuum. In fact, according to \eqref{bounce}, the Gauss-Bonnet term does not affect the behaviour of pole. In other words, if $X \approx 1$, in the absence of the Gauss-Bonnet term, $B_{M_+ \to AdS_-}$ still has the same behavior. We can give an intuitive understanding for this result. According to \eqref{Lorentzianaction}, the GR term is proportional to curvature $R$ (or $k^2$) while the Gauss-Bonnet term is proportional to $R^2$ (or $k^4$). Therefore, when $k$ goes to zero, the GR term should play the dominant role in tunnelling exponent and the GB term cannot affect the behavior near $k \approx 0$. 

We have a comment on $X=1$. If $X=1$, even if there is no Gauss-Bonnet term, according to \eqref{XDnoGB}, the precise value $X$ is dependent on the membrane tension and the depth of the daughter vacuum. In any given model, these quantities are exposed to radiative corrections. We need to tune the value of $X$ to be close to unity, but this is not natural. If we include the contribution from the Gauss-Bonnet term, we can expect that more quantities should be tuned. Therefore, we will not accept $X=1$ limit. 

There is an argument in \cite{BD2012} that up-tunneling is impossible from Minkowski or anti-dS space. The argument is the following. Every instanton describes not one but two transitions. If we calculate the rate for one process, then we should calculate the rate for the reverse process as well. In other words, this means that the instanton not only describes “down-tunneling” from A to B, but also describes “up-tunneling” from B to A \cite{BD2012}. These processes are described by the same instanton both, and the rate to down-tunneling and up-tunneling are given by
\begin{equation} \label{downtunnel}
    \Gamma_{A \rightarrow B} \sim e^{-\Delta S_E/ \hbar}, \quad \Delta S_E = S_E(\text{instanton})-S_E(A),
\end{equation}
and 
\begin{equation} \label{uptunnel}
    \Gamma_{B \rightarrow A} \sim e^{-\Delta S_E/ \hbar}, \quad \Delta S_E = S_E(\text{instanton})-S_E(B).
\end{equation}
The ratio of \eqref{downtunnel} and \eqref{uptunnel} is given by $e^{S_E(A)-S_E(B)}$, the ratio of the exponentials of the entropies. The essential feature of de Sitter space which allows up-tunnelling is that the system has finite size in the sense that it has a finite horizon volume and a finite entropy, namely, $S_E(B)$ is finite. However, when sizes of system are infinite, entropy differences between two states are infinite, therefore based on principle of detailed balance the up-tunneling is impossible. Anti-de Sitter space and Minkowski space have infinite horizon volumes and infinite entropies, therefore in \cite{BD2012} the authors claimed that up-tunnelling is impossible from anti-de Sitter space or Minkowski space. So the readers may ask the question: do our results contradict the argument in \cite{BD2012}? 

Here is our explanation. As we have seen in Table 2, we have a new decay branch $AdS_{+} \rightarrow dS_{-}$, when $\epsilon = \pm 1$. In fact, the ratio of \eqref{downtunnel} and \eqref{uptunnel} is given by $e^{S_E(dS)-S_E(AdS)}$. Since $S_E(dS)$ is finite while $S_E(AdS)$ is infinite, the difference between these two action is infinite. However, this does not mean that $B_{AdS_{+} \rightarrow dS_{-}}$ needs to be infinite. In General Relativity and other previous models \cite{us,BD2012}, the decay branch $B_{AdS_{+} \rightarrow dS_{-}}$ does not exist because of the negative membrane tension. However, as we have pointed out, when we include the contribution of the Gauss-Bonnet term, there exists a new decay branch $B_{AdS_{+} \rightarrow dS_{-}}$, when $\epsilon = \pm 1$, which has positive membrane tension. Hence there is no contradiction between our result and the argument of \cite{BD2012}.

\section{Discussions} \label{sec:condis}
The cosmological constant problem is a long-standing problem in theoretical physics. In the current string cosmology, anthropic principle is the dominant explanation for the small value of the cosmological constant of our universe. We challenged the dominant assertion in our previous paper \cite{us}. For a family of string-inspired models of four-form fluxes coupled to scalar fields, including Bousso and Polchinski's set-up, we can show how the current vacuum can be selected on “probabilistic” grounds. In other words, our universe is born from probability. Hence the existence of our world is independent of mankind and other biological species.  
 
The main idea of our paper \cite{us} was inspired by the model constructed by Kaloper \cite{Kaloper:2022oqv} and  Kaloper and Westphal \cite{Kaloper:2022jpv}, though we have shown how to generalize this special model. We proposed an important parameter $X$ which controls the lifetime of low scale vacua, similar to the current vacuum we are living. When the parent vacuum goes to Minkowski spacetime, we found that a pole can appear in the corresponding bounce if $X$ is suitably bounded. This can guarantee the stability of the Minkowski vacuum and hence the longevity of those vacua near Minkowski in Planck units.  

In paper \cite{us}, we only considered the cosmological constant problem in $4D$. If we are interested in the properties of string landscape in higher dimensions, we can ask that if the tunneling exponent $B$ has a similar behavior in arbitrary $D$-dimension as in $4D$. Moreover, the Gauss-Bonnet term appears often in string theory and play an important role in gravitational theory \cite{CELP2022}. So we can consider transition between vacua in $D$-dimension including the Gauss-Bonnet term. According to higher dimensional theory motivated by string theory, the associated coupling constant of the Gauss-Bonnet term $\theta$ should be positive, otherwise the whole theory will be unstable \cite{CP2008}. In the previous calculation, we assume that $\theta$ is a positive constant since $\phi_i$ are locally constants. We find that the tunneling bounce $B$ of $\text{Min}_+ \to \text{AdS}_-$ is proportional to $k^{-(D-2)}_{+}$, which has a pole at $k^2_{+}$ and is similar to the result in $4D$. This means that the Gauss-Bonnet term does not play a major role near the pole. Furthermore, we should point out that when we include the Gauss-Bonnet term, there exist two possibilities if we require positive membrane tension. As a result, we can find a new decay branch $AdS_{+} \rightarrow dS_{-}, \ \epsilon_{\pm} = 1$. This new branch does not exist in General Relativity and our previous model \cite{us}. 

Although the Gauss-Bonnet term is a topological term when $D=2,3,4$, we can still consider the $4D$ limit, i.e., $D \rightarrow 4$ Gauss Bonnet gravity \cite{GL2020}. In the strong coupling limit ($\theta \rightarrow \infty$), the Gauss-Bonnet term in $4D$ gives rise to non-trivial contributions to gravitational dynamics, while preserving the number of graviton degrees of freedom and being free from Ostrogradsky instability \cite{GL2020}. However, so far the up-tunneling in $D \rightarrow 4$ Gauss Bonnet gravity has not been found. We need to clarify two points. First, although it is not central to the results of the current paper, the strong regime limit as $D \rightarrow 4$ is still unclear and has been critiqued by several authors, such as \cite{hennigar2020taking}. Second, the inclusion of the GB term is not without problems. As demonstrated in \cite{reall2014causality}, the equations are not weakly hyperbolic when the spacetime curvature becomes sufficiently large. 

Based on our results, there are at least two speculative pictures. One picture is that AdS vacua decay to near Minkowski vacua ($k^2_{-}>0$) directly. We can expect that there is a pole $k^{-2}_{-}$ near Minkowski vacua. The other possible picture is that in the first step vacua decay from $AdS$ parent vacua, which can be obtained in string theory, to high $dS$ daughter vacua. And then in the second step high $dS$ vacua decay to near Minkowski vacua $(k^2_{-}>0)$. In order to get a very small positive cosmological constant, we need to introduce a mechanism to halt the decay process. The $dS$ vacua cannot be long-lived. In fact it is very hard to get de Sitter vacua from string theory \cite{nods}. Many scenarios have been tried, such as KKLT scenario \cite{KKLT2003}. We should find more terms which can induce $AdS \to dS$ decay channel, which is new and interesting. At present we have not figured out the meaning of the new decay branch in string landscape. All of these should be studied in the future.

\paragraph{Acknowledgements} 
YL was supported by an STFC studentship. Thanks for the discussion with Antonio Padilla, Francisco G. Pedro, Benjamin Muntz and Trevor Cheung. For the purpose of open access, the authors have applied a CC BY public copyright licence to any Author Accepted Manuscript version arising.

\appendix

\section{Calculations of the Gauss-Bonnet term in $D$-dimension}
We consider $O(D)$ symmetric Euclidean field configurations, with metric
\be \label{metric}
ds^2=dr^2+\rho(r)^2 d \Omega_{D-1},
\ee
where $d \Omega_{D-1}=h_{ij} d\xi^i d\xi^j$ is the metric on a unit $(D-1)$-sphere. We denote $r$ as the radial coordinate and $i$'s as the other coordinate components. Then we can obtain the corresponding components of Riemann tensor
\be \label{Riet1}
R^i_{\ jkl}= \left[\frac{1}{\rho^2} - \left(\frac{\rho'}{\rho} \right)^2 \right] (\delta^i_{\ j} \delta_{kl}- \delta^i_{\ l} \delta_{jk}),
\ee
\be \label{Riet2}
R^i_{\ rrj}= \frac{\rho''}{\rho} \delta^i_{\ j},
\ee
and other components are vanishing. Then by contraction we have the components of Ricci tensor in $D$-dimension:
\be \label{Rict1}
R^i_{\ k}= \left( (D-2)\left[\frac{1}{\rho^2} - \left(\frac{\rho'}{\rho} \right)^2 \right] - \frac{\rho''}{\rho} \right) \delta^i_{\ k},
\ee
\be \label{Rict2}
R^i_{\ r}= 0,
\ee
\be \label{Rict3}
R^r_{\ r}= -(D-1)\frac{\rho''}{\rho}.
\ee
As a result, we can get the Ricci scalar in $D$-dimension:
\be \label{Rics}
R= (D-1)(D-2) \frac{1}{\rho^2} - (D-1)(D-2)\left(\frac{\rho'}{\rho} \right)^2 -2(D-1)\frac{\rho''}{\rho}.
\ee
Finally, the associated Gauss-Bonnet term is given by
\begin{eqnarray}\label{GB}
\begin{aligned}
R_{GB} \equiv & R^2 - 4 R^{\mu \nu} R_{\mu \nu} + R^{\mu\nu\rho\sigma} R_{\mu\nu\rho\sigma} \\
= &  (D-1)(D-2)(D-3)(D-4) \frac{1}{\rho^4} + (D-1)(D-2)(D-3)(D-4) \frac{\rho'^4}{\rho^4} \\
& - 2(D-1)(D-2)(D-3)(D-4) \frac{\rho'^2}{\rho^4} -4 (D-1)(D-2)(D-3) \frac{\rho''}{\rho^3} \\
& + 4 (D-1)(D-2)(D-3) \frac{\rho'^2 \rho''}{\rho^3}.
\end{aligned}
\end{eqnarray}

\section{Calculation of the boundary term in $D$-dimension}
The extrinsic curvature $K_{ij}$ of metric \eqref{metric} is
\be \label{ec}
K_{ij}= \frac{\rho'}{\rho} \gamma_{ij},
\ee
and its trace is
\be \label{ect}
K \equiv \gamma^{ij}K_{ij} = (D-1) \frac{\rho'}{\rho}.
\ee
Then according to \eqref{Jij}, then we have
\be \label{Jij2}
J_{ij} = -\frac{1}{3} (D-2)(D-3) \frac{\rho'^3}{\rho^3} \gamma_{ij},
\ee
and its trace is
\be \label{Jij3}
J \equiv \gamma^{ij}J_{ij} = -\frac{1}{3} (D-1)(D-2)(D-3) \frac{\rho'^3}{\rho^3}.
\ee
The corresponding Ricci tensor $\hat{G}_{ij}$ on the spacetime, $\Sigma$, is given by
\be \label{hGij}
\hat{G}_{ij} \equiv \hat{R}_{ij} - \frac{1}{2} \gamma_{ij} \hat{R} = (D-2) \frac{1}{\rho^2} \delta_{ij} - \frac{1}{2} \gamma_{ij} (D-1)(D-2) \frac{1}{\rho^2}.
\ee
Then we can obtain that
\be \label{J-2hGijKij}
J-2\hat{G}_{ij} K^{ij} = (D-1)(D-2)(D-3) \left(\frac{\rho'}{\rho^3}- \frac{1}{3} \frac{\rho'^3}{\rho^3} \right).
\ee
Finally, the first term of the boundary term \eqref{Lorentzianboundary}, the Gibbons-Hawking term, is given by
\be \label{GHt}
\int_{\Sigma} d^{D-1}x \sqrt{|\gamma|} M^{D-2}_{pl} K = \int_{\Sigma} d^{D-1}x M^{D-2}_{pl} (D-1) \frac{\rho'}{\rho},
\ee
and the second term of the boundary term \eqref{Lorentzianboundary}, the Myers term, is given by
\be \label{Myerst}
\int_{\Sigma} d^{D-1}x \sqrt{|\gamma|} 4\theta (J-2\hat{G}_{ij} K^{ij}) = 4\theta (D-1)(D-2)(D-3) \int_{\Sigma} d^{D-1}x \sqrt{|\gamma|} \left(\frac{\rho'}{\rho^3}- \frac{1}{3} \frac{\rho'^3}{\rho^3} \right).
\ee


\begin{thebibliography}{99}

\bibitem{us}
Y.~Liu, A.~Padilla and F.~G.~Pedro,
JHEP \textbf{10} (2023), 014
doi:10.1007/JHEP10(2023)014
[arXiv:2303.17723 [hep-th]].

\bibitem{LPP2024}
Y.~Liu, A.~Padilla and F.~G.~Pedro,
[arXiv:2404.02961[hep-th]].

\bibitem{Enz}
C.~P.~Enz, ~A.~ Thellung,
Helv. Phys. Acta 33 (1960) 839-848

\bibitem{Weinberg:1988cp}
S.~Weinberg,
Rev. Mod. Phys. \textbf{61} (1989), 1-23

\bibitem{Polchinski:2006gy}
J.~Polchinski,
[arXiv:hep-th/0603249 [hep-th]].

\bibitem{Burgess:2013ara}
C.~P.~Burgess,
[arXiv:1309.4133 [hep-th]].

\bibitem{Padilla:2015aaa}
A.~Padilla,
[arXiv:1502.05296 [hep-th]].

\bibitem{Kaloper:2022oqv}
N.~Kaloper,
Phys. Rev. D \textbf{106} (2022) no.6, 6
[arXiv:2202.06977 [hep-th]].

\bibitem{Kaloper:2022jpv}
N.~Kaloper and A.~Westphal,
Phys. Rev. D \textbf{106} (2022) no.10, L101701
[arXiv:2204.13124 [hep-th]].

\bibitem{Henneaux:1989zc}
M.~Henneaux and C.~Teitelboim,
Phys. Lett. B \textbf{222} (1989), 195-199

\bibitem{Padilla:2014yea}
A.~Padilla and I.~D.~Saltas,
Eur. Phys. J. C \textbf{75} (2015) no.11, 561
[arXiv:1409.3573 [gr-qc]].



\bibitem{Bousso:2000xa}
R.~Bousso and J.~Polchinski,
JHEP \textbf{06} (2000), 006
[arXiv:hep-th/0004134 [hep-th]].

\bibitem{Polchinski:1998rr}
J.~Polchinski,
String theory. Vol. 2: Superstring theory and beyond, 
Cambridge Monographs on Mathematical Physics, Cambridge University Press, 12, 2007.

\bibitem{CELP2022}
F.~Cunillera, W.T.~Emond, A.~Lehébeld and A.~Padilla,
JHEP \textbf{02} (2012), 012
doi:10.1007/JEHP02(2012)012
[arXiv:hep-th/2112.05771 [hep-th]].

\bibitem{GL2020}
D.~Glavan1 and C.~Lin,
Phys.Rev.Lett \textbf{124}(2020), 081301
doi:10.1103/PhysRevLett.124.081301
[arXiv:1905.03601 [gr-qc]].


\bibitem{CP2008}
C.~Charmousis, and A.~Padilla,
JHEP \textbf{12} (2008), 038
doi:10.1088/1126-6708/2008/12/038
[arXiv:hep-th/0807.2864 [hep-th]].


\bibitem{PS2012}
A. Padilla and V. Sivanesan,
JHEP \textbf{1208} (2012) 122
[arXiv:1206.1258 [gr-qc]].

\bibitem{Davis2003}
S.~C.~Davis,
Phys. Rev. D \textbf{67} (2003) 024030
[hep-th/0208205].

\bibitem{CLP2019}
B.~Coltman, Y.~Li and A.~Padilla,
JCAP \textbf{06} (2019) 017
[arXiv:1903.02829 [hep-th]].


\bibitem{Gibbons:1976ue}
G.~W.~Gibbons and S.~W.~Hawking,
Phys. Rev. D \textbf{15} (1977), 2752-2756


\bibitem{Duncan:1989ug}
M.~J.~Duncan and L.~G.~Jensen,
Nucl. Phys. B \textbf{336} (1990), 100-114



\bibitem{Coleman:1977py}
S.~R.~Coleman,
Phys. Rev. D \textbf{15} (1977), 2929-2936
[erratum: Phys. Rev. D \textbf{16} (1977), 1248]

\bibitem{Callan:1977pt}
C.~G.~Callan, Jr. and S.~R.~Coleman,
Phys. Rev. D \textbf{16} (1977), 1762-1768

\bibitem{Coleman:1980aw}
S.~R.~Coleman and F.~De Luccia,
Phys. Rev. D \textbf{21} (1980), 3305



\bibitem{BD2012}
A.~R.~Brown and A.~Dahlen,
Phys. Rev. D \textbf{85} (2012), 104026.

\bibitem{Giudice:2019iwl}
G.~F.~Giudice, A.~Kehagias and A.~Riotto,
JHEP \textbf{10} (2019), 199
[arXiv:1907.05370 [hep-ph]].

\bibitem{Kaloper:2019xfj}
N.~Kaloper and A.~Westphal,
Phys. Lett. B \textbf{808} (2020), 135616
[arXiv:1907.05837 [hep-th]].


\bibitem{Moretti:2022xlc}
M.~Moretti and F.~Gil Pedro,
JHEP \textbf{08} (2022), 287
[arXiv:2202.07004 [hep-th]].



\bibitem{nods}
Ulf H.Danielsson and Thomas Van Riet, 
Int.J.Mod.Phys.D \textbf{27} (2018) 12, 1830007
doi.org/10.1142/S0218271818300070
[arXiv:1804.01120 [hep-th]].


\bibitem{KKLT2003}
S.~Kachru, R.~Kallosh, A.~Linde1 and S.~P.~Trivedi,
Phys. Rev. D \textbf{68} (2003), 046005
doi:10.1103/PhysRevD.68.046005
[arXiv:0301240 [hep-th]].

\bibitem{hennigar2020taking}
Robie A. Hennigar, David Kubiznak, Robert B. Mann and Christopher Pollack,
JHEP \textbf{07} (2020),027
doi.org/10.1007/JHEP07(2020)027
arXiv:2004.09472

\bibitem{reall2014causality}
Harvey S. Reall, Norihiro Tanahashi and Benson Way,
Class. Quantum Grav., \textbf{31} (2014), 205005
doi.org/10.1088/0264-9381/31/20/205005
arXiv:1406.3379



\end{thebibliography}
\end{document}